\documentstyle[11pt,newpasp,twoside]{article}
\def\edcomment#1{\iffalse\marginpar{\raggedright\sl#1\/}\else\relax\fi}
\reversemarginpar

\input psfig.tex

\def\3h{{\scriptscriptstyle 3/2}}
\def\ie{{\frenchspacing \sl i.e.~}}
\def\eg{{\sl e.g.~}}
\def\cf{{\sl c.f.~}}
\def\viz{{\sl viz.~}}
\def\etal{{\sl et al.~}}
\def\loccit{{\sl loc. cit.~}}
\def\Teff{{T$_{ef\!f}$}}
 \def\ahh{{\scriptscriptstyle 1/2}}
 \def\trhh{{\scriptscriptstyle{3/2}}}

 \def\M{\displaystyle}
 \def\Y{{\cal Y}}
\def\th{\thinspace}
\def\ni{\noindent}
\def\at{{\rm\char'100}}

\begin{document}

\title{Nonlinear Pulsations of Convective Stellar Models}

\author{J.~Robert~Buchler{\altaffilmark{1}}}

\affil{Physics Department, University of Florida,
Gainesville, FL 32611, USA }

\altaffiltext{1}{e--mail: buchler\at phys.ufl.edu}

\begin{abstract}
 We review the numerical modelling of the nonlinear pulsations of classical
variable stars with hydrocodes that include the effects of turbulent
convection.  Despite their simplicity these turbulent convective recipes appear
to remove many of the difficulties that radiative codes faced.  In particular,
the numerical modelling of double mode pulsations has become possible.
\end{abstract}



\section{Introduction}

Richard Feynman is said to have once remarked something to the effect "now that
we have solved the proble of quantum electrodynamics, we will have to solve the
real hard problems such as how water flows in a pipe".  The stellar problem,
because of convection on top of turbulence and the compressibility of the
fluid, is even harder to tackle and several generations of astrophysicists have
tried to come to terms with this problem.  Turbulence and convection (TC) are
necessarily 3 dimensional phenomena, and with the development of faster
computers increasingly realistic numerical simulations are being made, although
it will be a long time before their spatial resolution approaches that required
by the large stellar Rayleigh and small Prandtl numbers.  In the meantime
stellar physicists continue to attempt to reduce TC to a 1D recipe and thus to
a mere subroutine that can be used in stellar evolution or pulsation
calculations (for recent update on astrophysical convection \cf \eg Buchler \&
Kandrup 2000).  In this paper we review some interesting recent developments in
nonlinear pulsation calculations.

The seminal and most influential work has been the mixing length theory (MLT)
of Erika B\"ohm-Vitense (\cf Cox \& Giuli 1968), and many of the newer recipes
are extensions of MLT.  In its original form it consists of an instantaneous,
local approximation in which the convective flux is proportional to the 3/2'th
power of the convectively unstable entropy gradient, $F_c\propto (-ds/dr)^\3h$.

The TC recipe that works best in stellar evolution is not necessarily the best
for stellar pulsations.  Indeed, in stellar evolution the convective timescales
are typically many orders of magnitude smaller than the evolutionary
time-scales.  Furthermore, convective overshooting is very important because it
mixes the chemical elements with often drastic consequences for nuclear burning
and the subsequent evolution.  In contrast, mixing plays no role in stellar
pulsation because the pulsating envelopes are chemically homogeneous.  But here
we have large velocity fields and shear motions, so that time-dependence of TC
may have to be taken into account.  Thus the pulsation timescales, while
generally longer than the convective time-scales, are sufficiently close so
that there can be a feedback between pulsation and convection.  This feedback
is further enhanced because the convective regions that are caused by large
opacity are also regions where pulsational driving occurs.  An illustration of
the time-dependence of the turbulent energy and convective flux during a
pulsational cycle has been presented in Buchler, Yecko, Koll\'ath, \& Goupil
(1999, [BYKG99], Figs.~1 and 2).

How good are MLT and its extensions?  An important 3D simulation by Cattaneo,
Brummell \& Toomre (1991) indicated first, that the convective flow is
dominated by large downflows, but that these flows are 'convectively neutral'
in the sense that they carry as much kinetic energy downward as enthalpy
upward, and second, that the convective flux is dominated by small scale
upflows, precisely the type of picture that underlies MLT.  However, for
computational reasons, the Prandtl numbers used in the calculations were orders
of magnitude larger than the stellar ones, and the Rayleigh numbers orders of
magnitude smaller.  Furthermore, the boundary conditions were fixed, whereas in
a star convection has to adjust itself so that together with radiation it
carries the given total energy flux (in a static context).

We recall the hydrodynamic equations in the context of radial stellar
pulsation:

\vskip -5pt
 \begin{eqnarray}
 {du\over dt} &=& -{1\over \rho} {\partial \over \partial r}
(p+{p_t} + {p_\nu}) -{GM\over r^2} \\
 {de\over dt} + p {dv\over dt} &=& -{1\over \rho r^2}
 {\partial\over \partial r} [r^2 (F_r+{F_c})] - {\cal C}.
 \end{eqnarray}

For the hydrodynamics all we need is a recipe for the turbulent pressure
$p_t$, the eddy viscous pressure $p_\nu$, the convective flux $F_c$ 
and the source and sink of turbulent energy ${\cal C}$.

\section{The Turbulent Convective Model Equations}

Many recipes have been suggested to compute these four quantities.  A very
nice, albeit dated, review is that of Baker (1987), and for an update see
Montesinos \etal (1999).  Many of these are far too complicated (\eg up to 10
nonlinear PDEs) and numerically tricky to implement in hydrocodes.  Since this
review concerns primarily stellar pulsations with an emphasis on nonlinear
calculations, we will limit ourselves to mentioning the time-dependent recipes
that have actually been used in nonlinear hydrodynamics calculations.  All
these recipes involve the addition of a single time-dependent diffusion
equation for the turbulent energy $e_t$

 \begin{equation}
 {de_t\over dt} + (p_t+p_\nu) {dv\over dt} = -{1\over \rho r^2}
{\partial\over \partial r} (r^2 F_t) + {\cal C}
 \end{equation}
The ancillary, defining equations are

 \begin{tabular}{lcll}
 \vspace{2mm}
 $\M {\cal C}$  &=& ${\cal S} - { \epsilon}$               &\\
 \vspace{2mm}
 $\M  \epsilon$ &=& $\alpha_d \th \th e_t^\trhh/ \Lambda $ &\\
 \vspace{2mm}
 $\M \Lambda$   &=& $\alpha_\Lambda H_p $                  &\\
 \vspace{2mm}
 $\M H_p$       &=& $\M (d\th \ln p/dr)^{\scriptstyle -1} = p/(\rho g) $ &\\
 \vspace{2mm}
 $\M {  p_t}$   &=& $\M\alpha_p \th \rho \th e_t$          &\\
 \vspace{2mm}
 $\M {  p_\nu}$ &=& $-\M\alpha_\nu\th\Lambda\th \rho \th e_t^\ahh\th
  r (\partial (u/r)/\partial r)$                           &\\
 \vspace{2mm}
 $\M {  F_t}$   &=& $-\M\alpha_t\th\alpha_\Lambda\th \rho H_p\th
    e_t^\ahh\th  (\partial e_t/\partial r)$                &\hskip 5.5cm (4)
\end{tabular}
\vskip 10pt

For the sake of simplicity these equations disregard some features of
convection and have therefore their shortcomings.  They are based on a
diffusion approximation ($F_c \propto ds/dr$ and $F_t \propto de_t/dr$) and
ignore nondiffusive transport, \eg by plumes.  They also disregard pressure
fluctuations and are limited to subsonic convective velocities.  Radiative
losses in the convection however can easily be incorporated (Wuchterl \&
Feuchtinger 1998, Buchler \& Koll\'ath 2000 [BK00]).  It is only a comparison
with observational constraints that will ultimately decide on the quality of
the approximations in the context of stellar pulsations.

The recipes that have been used in the nonlinear codes fall into three
groups depending on the choice of the functional relationships of $F_c$
and {\cal C} on $e_t$ and the dimensionless entropy gradient which we
call $\Y\equiv (H_p/c_p)\th ds/dr$ (not to be confused with the helium
abundance).  We refer to BK00 for further
details.  They are the (1.) Stellingwerf (1982) [S] formulation, used
by the Italian group (Bono \& Stellingwerf, 1994, Bono, Caputo,
Castellani \& Marconi 1997 [BCCM97]) and by Gehmeyr (1992), (2.) the
Kuhfu\ss [K] formulation (1986, \cf also Gehmeyr \& Winkler 1992) used
in the Vienna code (Feuchtinger 1998, Feuchtinger 1999a [F99a])
and (3) a hybrid Florida [FL] formulation (Yecko, Koll\'ath, Buchler 1998,
[YKB98]), that has been used by Koll\'ath, Beaulieu, Buchler \& Yecko 1998
[KBBY98]).  The Florida hydrocode has recently been extended to run with
all three schemes (see below), and, importantly, also to perform a {\sl
linear} stability analysis (linear periods and growth-rates).

 \vskip 10pt

 \begin{tabular}{lcllrr}
 \vspace{1mm}
 $F_c$       & = & $A/B$                  & $e_t\th (\Y)^\ahh$ & &   \\ 
 \vspace{3mm}
 ${ \cal S}$ & = & $\alpha_d B /\Lambda$  & $e_t\th (\Y)^\ahh$ &\hskip5cm(S)&\\
 \vspace{1mm}
 ${ F_c}$    & = & $A$                    & $e_t^\ahh \Y$      &  &  \\ 
 \vspace{3mm}
 ${ \cal S}$ & = & $\alpha_d B^2/\Lambda$ & $e_t^\ahh \th \Y$  & (KGW)&\\
 \vspace{1mm}
 ${ F_c}$    & = & $A$                    & $ e_t^\ahh\th\Y$   & &\\ 
 \vspace{3mm}
 ${ \cal S}$ & = &  $\alpha_d B /\Lambda$ & $e_t\th (\Y)^\ahh$ & (YKB)&
                                                             \hskip 0.8cm (5)\\
 \end{tabular}

\ni where
\vskip 5pt
 \begin{tabular}{lcll}
 $\M A$ & = &$\M\alpha_c \th \alpha_\Lambda\th \rho c_p T$
  \vspace{3pt} &\\
 $\M B$ & = &$\M \alpha_s \th \alpha_\Lambda \th
 \sqrt{p \beta T/\rho}
  = \alpha_s \th \alpha_\Lambda \th \sqrt{c_p T \nabla_{\!\!a}}
  = \alpha_s\alpha_\Lambda \sqrt{{\beta T/ \Gamma_1}} \th\th c_s $
 \vspace{3pt} &\\
 $\M \beta$ &=& $\M (\partial \ln v / \partial T)_p$ & \hskip 1.2cm (6)
 \end{tabular}
\vskip 5pt

The recipes involve a total of 7 dimensionless $\alpha$ parameters that
are of order unity, but for which theory gives little guidance.  They
ultimately have to be calibrated by comparing the numerical results to
the stellar observations.

The three schemes have been compared in BK00.  In the stationary limit and with
the disregard of overshooting (no $F_t$ eq.~3 reduces to ${\cal C}=0$.  In
other words, it gives an expression for $e_t$ in terms of the local,
instantaneous physical quantities such as density and temperature and their
gradients, and the ancillary equation~6 provides the convective flux.  This is
then equivalent to standard MLT.  However, in standard MLT the $p_t$ and
$p_\nu$ are omitted (which is not a good approximation; \cf below), although
they could be readily included once $e_t$ is known.
 
In the time-dependent context one might expect the three formulations to have a
very different behavior.  However, as shown in BK00 the growth-rates differ
very little, and furthermore the three recipes give essentially the same limit
cycles as well (Figs.~3 and 6 of BK00).

From these, albeit limited comparisons, one is tempted to conclude that most of
the differences between the published nonlinear results have more to do with
the choice of the $\alpha$ parameters\footnote{and some modifications such as
flux limiters (Wuchterl \& Feuchtinger 1998), small P\'eclet number corrections
(BK00), or sonic dissipation (Gehmeyer \& Winkler 1992).}  than with the choice
of the time-dependent diffusion equation for the turbulent energy.

\psfig{figure=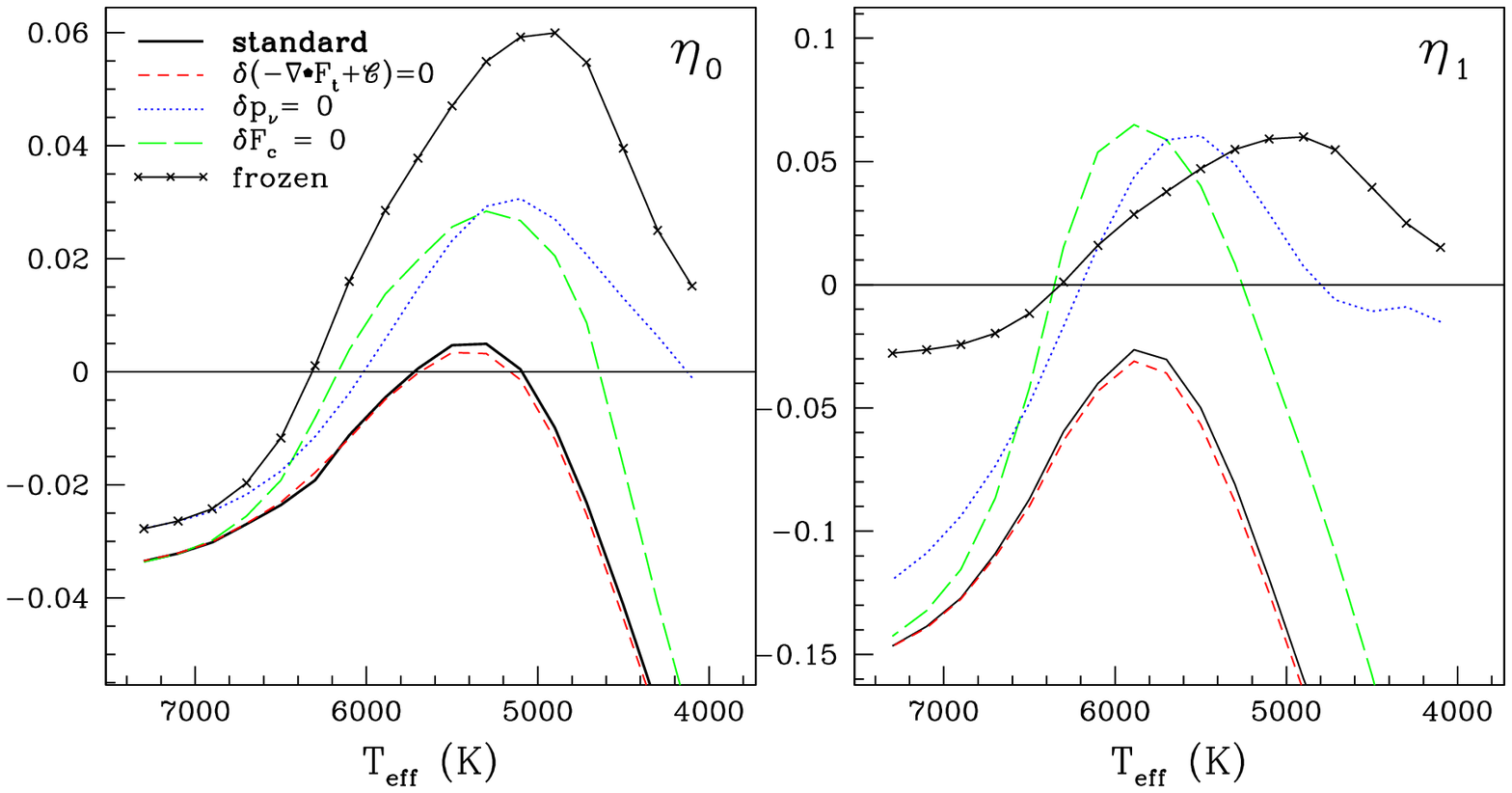,width=13.7cm}
\ni{\small Fig.~1: Linear stability of sequence of models. 
Comparison of different
approximations with exact results. Left: fundamental mode, right: first
overtone mode.}
\vskip 10pt

\section{Linear stability properties}

In evolutionary computations it is often customary to compute also the linear
stability of the evolving models to delineate the instability strips.  For
expediency several approximations are often made.  We want to stress again
(YKB98, BYKG98) that some of these approximations are not very good, as we go
on to show in Figure~1.  From right to left we display the normalized
growth-rates $\eta= 2 \kappa P$ of the fundamental and first overtone,
respectively, for a sequence of Cepheid models (M=5M$_{\rm \odot}$,
L=2060L$_{\rm \odot}$, Z=0.02) as a function of \Teff\. ($\kappa$ is the
growth-rate and P the period.)

The solid lines denote the nonadiabatic growth-rates, consistent with the hydro
and TC equations.  The crossed solid lines correspond to the frequently used
{\sl frozen flux approximation}: the MLT flux is included in the computation of
the static equilibrium model, but it is ignored in the computation of the
linear eigenvalues, and the eddy viscous pressure is also disregarded.  Clearly
this approximation misses even the blue edge by a $\approx$600K degrees for the
F mode and indicates instability for the O1 mode which is solidly stable.  In
the next two approximations everything is linearized correctly except (1) the
perturbation of the turbulent pressure is disregarded (dotted line); (2) the
perturbation of the convective flux is disregarded (long-dashed line) --
neither of these a good approximation.  The best 'simple' approximation is
(dashed line): MLT expression for the flux (derived from ${\cal C}-\nabla\cdot
F_t =0$), its linearization and inclusion of the eddy viscous pressure.  Of
course it is also important to choose 'proper' values of the $\alpha$
parameters.  A survey of the model behavior makes it quite clear that both a
convective flux {\sl and} a turbulent pressure are needed if one wants to get a
reasonable IS.  This importance of the eddy viscous pressure was already
pointed out a dozen years ago by Baker (1987).

In addition it should be remembered that nonlinear effects can shift the
linear IS boundaries by a few hundred degrees (\cf Fig.~2 below).

\section{RR Lyrae Models, RRab and RRc}

There have been several recent large surveys of nonlinear pulsations of single
mode RR Lyrae, both fundamental and first overtone pulsations by Bono \etal
[BCCM97] and by Feuchtinger (1999b) [F99b].  F99b also compares these results to
each other and to the available observations.  He finds that the light-curve
Fourier decomposition coefficients of both calculations agree faily well with
observations. but that there are discrepancies both between the two
calculations, and with observations in the low temperature models, \ie in the
most convective models.  The pulsation amplitudes of both calculations agree
well with observations (It should be noted though that this is not as stringent
a test as the Fourier parameters).

As far as the shapes of the lightcurves are concerned, BCCM97 obtain sharp, but
unobserved spikes (cf. their Figs.~2 and 4).  F99b (\cf also Wuchterl \&
Feuchtinger 1998) shows that these spikes are due to the fact that the
convective flux becomes larger than its physically allowed upper limit, \viz
$F_c < \rho c_p T u_{conv}$.  This is a result of the breakdown of the
diffusion approximation that is inherent in the TC equations.  They propose the
introduction of a 'flux limiter' to prevent the flux from exceeding this upper
limit.  As a result F99b obtains light curves that look much more like the
observed ones.

The RR Lyrae light-curves obtained by the Florida group do not have any
unphysical spikes either, despite the fact that no flux limiter has been
used.  Feuchtinger (private communication) has traced the absence of
spikes to a different choice of $\alpha$ parameters, those of the FL
group giving rise to lower overall turbulent energies and velocities.

The computed radial velocity curves do not agree very well with the
observations (F99b).  It seems that a more thorough calibration of the
$\alpha$ parameters is required before the final word is in on whether
such a simple 1D TC equation is capable of capturing the essence of
convection.

We also note that there are further constraints on the $\alpha$'s that
need to be taken into account.  For example, the observed temperature
(color) variations can be into account. Second, the double-mode RR Lyrae (RRd)
impose a number of sensitive additional constraints, namely the range of
periods and of temperatures over which they occur, as well as the values
of the component amplitudes.  Finally, we recall that the very frequent
Blazhko amplitude modulations have not yet been satisfactorily
explained, but that they are likely to also add constraints.

We note {\sl a propos} Blazhko effect that a possible mechanism for this effect
could be an interaction between pulsation and convection.  This becomes
particularly favored if normally real and stable convective (diffusion) modes
become oscillatory.  If these additional vibrational modes are only mildly
stable, and if their frequencies are in a $n:1$ resonance with the excited
pulsational mode, then a resonance condition could cause the convective mode to
interact nonlinearly with the pulsation and lead to amplitude modulations.  We
have checked on realistic stellar models that convective modes can indeed
become oscillatory, but we had to increase the time-scale for convection by a
very large factor, and no nonlinear computations have yet been performed.

\section{Morphology of the Instability Strip -- Modal Selection}

On the observational side the microlensing projects have produced global
pictures of the Cepheid instability strips for the SMC that are
absolutely stunning (Udalski \etal 1999, Beaulieu \etal 1995).

On the theoretical side, the turbulent convective hydrodynamics codes
have shed new light on the problem of modal selection in both Cepheids
and RR Lyrae stars.  Simultaneously, but totally independently,
Feuchtinger (1998) and KBBY98 found RR Lyrae models and Cepheid models,
respectively, that pulsated in the fundamental and first overtone modes
simultaneously, with stable, and steady amplitudes (\ie the models were
NOT switching modes).

On the basis of these computations (\cf BYKG 99 and Koll\'ath \etal in this
Volume) we can infer the following (schematic) Cepheid instability strips (IS)
in an HR diagram as shown in Fig.~2.  The left subfigure depicts a linear IS.
The first overtone IS in the form of a sugarloaf becomes stable above a certain
luminosity (and mass).  For simplicity we have omitted the second overtone from
the picture.

Of course, nonlinear effects change the domain in which the corresponding limit
cycles are stable.  The right-hand subfigure depicts a schematic of the
nonlinear Cepheid instability diagram.  The lines from the leftside subfigure
that are shown as solid lines are the blue and red boundaries of the
fundamental (F) and first overtone (O) IS's. (The now irrelevant parts of the
previously shown linear edges are shown as thin dotted lines.)  The new
additional solid lines are the nonlinear blue and red edges for the fundamental
and first overtone modes.  Thus the overtone red edge is shifted
somewhat to the left, but the fundamental blue edge can be shifted
substantially to the right.

Double-mode behavior occurs in the lower wedge-shaped region, delineated on the
left by overtone red edge and on the right by a dashed line.  Either
fundamental or first overtone (F/O) behavior occurs in the higher luminosity
region that is shown as dotted.  There may be a narrow region at the interface
of the DM and F regimes in which either DM or fundamental behavior can
occur.\footnote{In some computations, with different $\alpha$ parameters,
slightly more complicated interfaces have also been obtained .}  We note that a
good global understanding of all these regimes can be obtained with the help of
the amplitude equation formalism (\eg BYKG99).

 \vskip 15pt

\psfig{figure=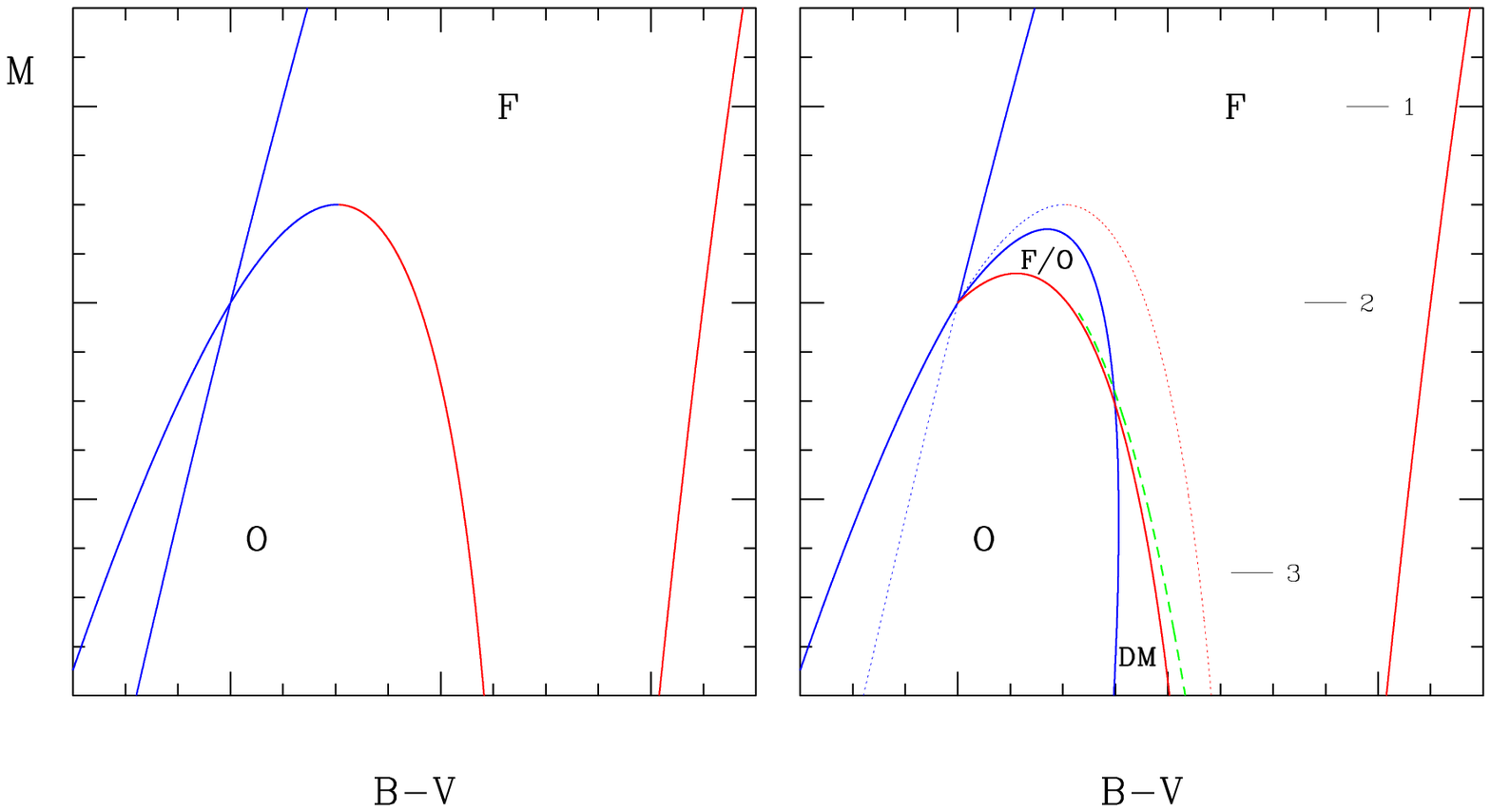,width=13.5cm}
{\noindent \small Fig.~2: 
Schematic Cepheid instability strip, {\it left}: linear,
{\it right}: nonlinear; \cf text.}

 \vskip 10pt

More specifically, for example at level 1 (high luminosity) in Fig.~2 only F
pulsations can occur.  At level 2 we have, going from high to low \Teff, a
regime first overtone pulsations, then a regime of either O or F pulsations
(hysteresis), and then F only pulsations.  At low luminosities, level 3, there
is first a regime of O only, then of DM only, with possibly a narrow regime of
either DM or F, followed by F only pulsations.

The hydrodynamic calculations indicate that DM behavior occurs only at
luminosities that can be noticeably lower than the tip of the overtone
instability strip.  This is in agreement with the SMC observations (Fig.~5 of
Udalski \etal 1999) which show a higher luminosity regime in which both F and O
Cepheids occur, and a lower luminosity one in which the DM Cepheids lie (with
the exception of a single star).

\vskip 5pt

The RR Lyrae stars, at least within a given cluster, have essentially
the same luminosity, mass and composition. The modal selection diagram
is therefore essentially the same as for a narrow horizontal strip in
the lower part of Fig.~2.

\section{Classical Cepheid Pulsations}

The classical Cepheids are much more diverse than the RR Lyrae stars.
They span a wide range of masses, luminosities and metallicities.
There are a great deal more observational constraints as well many of
which have been summarized in BK00.

For example, the overtone Cepheids have a maximum period $P_1^{max}$ which
occurs at the high luminosity tip of the IS.  Next, resonances play an
important role, \viz a $P_2/P_0=1/2$ resonance at about 10 days and a
$P_4/P_1=1/2$ resonance around 3--4.5 days.  This is evidenced by the structure
of the Fourier decomposition coefficients of the light and radial velocity
curves.  We note that in principle it is possible to obtain a purely
'pulsational' mass--luminosity relation by taking advantage of these two
resonances.  The periods and \Teff\ at which DM behavior can occur, as well as
the F/O, respectively O$_2$/O$_1$ amplitude ratios add a very tight constraint
as well.

 \vskip 15pt
\psfig{figure=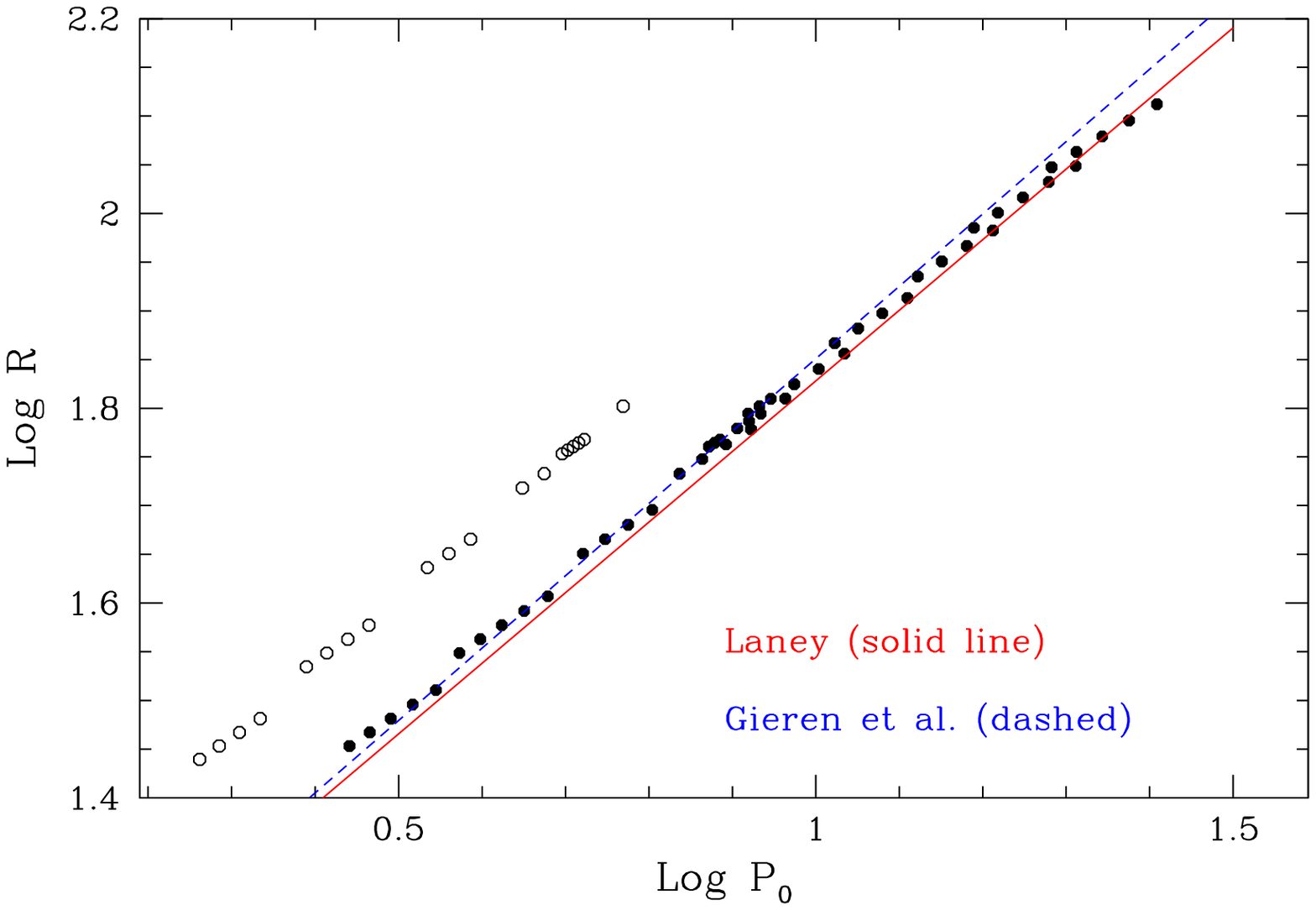,width=12.5cm}
{\noindent \small Fig.~3: Period-radius relation for Galactic Cepheids}
 \vskip 10pt

Radiative Cepheid models were found wanting in many respects, besides the
obvious one of not providing a red edge (for a review \cf Buchler 1998).  In
particular the discrepancies are largest for the low Z models for which the
linear growth-rates and consequently the pulsation amplitudes are much too
large.  The resonance masses are also much too small to agree
with stellar evolution calculations as was discussed in Buchler, Koll\'ath,
Beaulieu \& Goupil (1996).  The question therefore arises whether convection
can provide a differentially stronger dissipation for the low Z models.

There have been a number of Cepheid computations by the Italian, Vienna and
Florida groups, but no comprehensive calculations have yet been made to see if
all observational constraints can be simultaneously satisfied.  However, there
are some good news.  For example, with the convective hydrocodes it seems
possible to improve the light and radial velocity curves, and in particular to
obtain the wide excursion in the observed $\phi_{21}$ Fourier phase of the
overtone Cepheids in which purely radiative models had failed.  The most
dramatic achievement though is the modelling of DM behavior in Beat Cepheids
(KBBY98).

Interestingly, despite seven adjustable $\alpha$ parameters it was not possible
to impose both the observed upper limit for the period of the first overtone
Cepheids \underbar{and} obtain a reasonable width of the fundamental
instability strip!  This problem was solved when we included the physically
required correction for inefficient convection (small P\'eclet number) (BK00),
but at the expense of an additional, eighth free $\alpha$ parameter.

Some properties, such as period-radius relations seem relatively insensitive to
the values of the alphas.  In Figure~3 we show the P-R relation that we
obtain for Galactic Cepheid models, compared to the observational data.
A very similar agreement has been obtained by Bono \etal (1999).

Observations show that the SMC, LMC and Galactic Cepheids are remarkably
similar.  For example, the maximum first overtone periods lie around 6
days.\footnote{If one disregards V440 Per which may be an oddball.}.  They have
approximately the same luminosity, the same pulsation amplitudes, the same
Fourier decomposition coefficients, and the dominant F and O1 resonances are
almost in the same place, \ie near 10d and 3-4.5d, respectively.

The not-so-good news is that at this time it does not seem possible to obtain
good models both for the Galactic and for the low Z Magellanic Cloud Cepheids
with the same calibration of the eight $\alpha$ parameters.  Turbulent
convection does not provide larger dissipation for the low Z models.  The
difficulty that was encountered with the radiative models thus persists with
the convective models, and one may wonder whether the difficulty still lies
with the opacities, this time with H, He or with the lower temperature H$^-$
and molecular opacities.

 \section{Pop. II Cepheid Pulsations}

Pop. II Cepheids are variable stars that have lower metallicity than
thir classical siblings.  They also are believed to have much smaller
masses for the same luminosity, which makes them on the one hand much
more linearly unstable to vibrations, and on the other hand causes much
larger pulsation amplitudes.

Observationally these stars, known as W Vir and RV Tau stars, range from
periodic at low periods to strongly irregular at cycling times of 70 days.  The
irregular behavior seems to set in at period of about 25--30 days (Arp 1955,
Pollard, this volume).

The recent nonlinear analysis of the AAVSO observational data of R~Sct
(Buchler, Koll\'ath, Serre \& Mattei 1995) and of AC~Her (Koll\'ath, Buchler,
Serre \& Mattei 1998) showed very clearly that the mechanism for the irregular
behavior is the nonlinear interaction between the excited (linearly unstable)
mode and a (linearly stable) overtone.  Technically speaking, the dynamics
takes place in a 4D subspace of phase-space and the pulsations are thus
low-dimensional chaos,

This analysis corroborates numerical hydrodynamical results obtained a decade
ago which showed that the irregular behavior of W Vir models was also due to
low-dimensional chaos (Buchler \& Kov\'acs 1986, Kov\'acs \& Buchler 1997,
Aikawa 1987, Buchler, Goupil \& Kov\'acs 1987).  However, the onset of the
irregular behavior occurred in these radiative models with periods as low 8
days, \ie much lower than observations indicate.  Glasner \& Buchler (1990)
included a very simplistic MLT model in the hydro-code, and this pushed the
onset of chaos to higher periods.  More recent calculations with the TC Florida
code also show a shift in the same direction.

The basic nature of the irregular behavior is now understood,
but clearly more work is necessary to obtain more detailed agreement with
the observations.

\section{Mira Pulsations}

Convection plays an essential role in the cool and very extended Mira
variables, and they are hard to model with much confidence.  There is still a
debate about whether the stars pulsate in the fundamental or the first overtone
mode.

Ya'ari \& Tuchman (1996) have modelled the nonlinear pulsations of these stars
with very interesting results (see also Dorfi \& Feuchtinger in this Volume).
The large amplitude pulsations that develop cause a structural rearrangement of
the star.  Consequently the nonlinear period is quite different from both the
linear fundamental and first overtone periods.  However, convection is treated
with a standard time-independent MLT approach, and unfortunately eddy viscosity
is ignored in their computations.  The latter reduces the pulsational
amplitude, and could cause a qualitative change in the results.

\section{Conclusions}

In recent years several groups have included a description of turbulent
convection in their hydrocodes.  The addition of a simple nonlinear
time-dependent diffusion equation for turbulent energy with concomitant
convective flux and eddy viscous pressure has led to important improvements in
RR Lyrae and in Cepheid models.  Most striking has been the ability of these
codes to model DM pulsations in both RR Lyrae and Cepheids.

However, it is clear that small discrepancies remain in the RR Lyrae models,
both the the single mode RRab and RRc, as well as in the double-mode RRd.  The
next step is to see if a better calibration of the free $\alpha$ parameters can
bring us in a better agreement with the plethora of observational data.

In the Cepheid modelling, one obtains reasonable agreement for the Galactic
Cepheids, even with a preliminary crude calibration, but for the time being it
remains a puzzle why the low Z models fail so strikingly.

\section{Acknowledgements}

It is a pleasure to acknowledge numerous valuable discussions with my
collaborators Z. Koll\'ath, P. Yecko, M.-J. Goupil, J.-P. Beaulieu and
M. Feuchtinger. This work has been supported by the National Science
Foundation (AST9528338, AST9819608, INT9820805).

 \end{document}